\documentclass[12pt]{article}
\usepackage{epsf}
\title{ {\bf
The effects of non-universal extra dimensions on the
$\tau\rightarrow \mu \, \bar{\nu_i} \, \nu_i$ decay}}

\author{\vspace{1cm}\\
        {\bf E. O. Iltan}
        \thanks{E-mail address:
        eiltan@heraklit.physics.metu.edu.tr}
 \\
        Physics Department, Middle East Technical University \\
        Ankara, Turkey\\}

\date{}

\begin{document}
\setlength{\baselineskip}{24pt}
\maketitle
\setlength{\baselineskip}{7mm}
\begin{abstract}
We study $\tau\rightarrow \mu \,\bar{\nu_i}\, \nu_i$,
$i=e,\mu,\tau$ decays in the two Higgs doublet model, with the
inclusion of one and two spatial non-universal extra dimensions.
We observe that the branching ratio is sensitive to two extra
dimensions in contrary to a single extra dimension.
\end{abstract}
\thispagestyle{empty}
\newpage
\setcounter{page}{1}
\section{Introduction}
Lepton flavor violating (LFV) interactions are rich to analyze
since they exist in the loop level and they are sensitive the
physics beyond the standard model (SM). On the other hand they are
clean theoretically because they are free from the nonperturbative
effects. The experimental work done and the numerical results
obtained stimulate the theoretical studies on these decays. The
$\mu\rightarrow e\gamma$ and $\tau\rightarrow \mu \gamma$
processes are among the LFV decays and the experimental the
current limits for their branching ratios (BR) have been predicted
as $1.2\times 10^{-11}$ \cite{Brooks} and $1.1\times 10^{-6}$
\cite{Ahmed}, respectively.

There is an extensive theoretical work done on the LFV decays in
the literature. Such interactions are studied in a model
independent way in \cite{Chang}, in the seesaw model
\cite{Ilakovac}, in the framework of the two Higgs doublet model
(2HDM) \cite{ilt, iltHay}, in supersymmetric models
\cite{Barbieri1,
Barbieri2,Barbieri3,Ciafaloni,Duong,Couture,Okada}.

In the present work we study $\tau\rightarrow \mu \,\bar{\nu_i}\,
\nu_i$, $i=e,\mu,\tau$ decay in the model version of the 2HDM with
the inclusion of the non-universal extra dimensions. This process
exists at least at one loop level in the model III. The lepton
flavor violation is driven by the internal scalar bosons $h^0$ and
$A^0$ and the transition $\tau\rightarrow \mu$ is obtained.
Furthermore the internal Z boson connects this transition to the
$\bar{\nu}\nu$ output (see Fig. \ref{fig1ver}). Notice that we
respect the assumption of the non-existence of
Cabibbo-Kobayashi-Maskawa (CKM) type matrix in the leptonic sector
and vanishing charged Flavor Changing (FC) interactions.

With the inclusion of additional dimensions there may be an
enhancement in the BR of the decay under consideration. The extra
dimensions have been studied in the literature extensively
\cite{Dvali}-\cite{iltanEDM}. These new dimensions could not be
detected at present and the most favorable description is their
compactification on a surface with small radii. From the 4D point
of view this compactification results in Kaluza-Klein (KK) modes
of the particles with masses regulated by the parameter $R$, which
is a typical size of the extra dimension. If all the fields are
accessible to the extra dimensions,  the extra dimensional
momentum, and therefore, the KK number at each vertex is
conserved. Such extra dimensions are called universal extra
dimensions (UED) and they are studied in various works
\cite{ArkaniHamed, Arkani,
Antoniadis1,Antoniadis3,Papavassiliou,Chakraverty,Agashe,iltmuegam}
in the literature. If the extra dimensions are accessible to some
fields but not all in the theory, such type of extra dimensions
are called non-universal extra dimensions (NUED) (see for example
\cite{Dienes, Agulia, Ruckl, iltanEDM}). In this case the KK
number at each vertex is not conserved and tree level interaction
of KK modes with the ordinary particles can exist.

In our work we take the extra dimensions as non-universal and
assume that the first Higgs doublet and the gauge fields are
accessible to the extra dimensions, however the leptons, the
quarks and the second Higgs doublet, which contains new Higgs
particles, live in the 4D brane. In this case the contributions
due to the additional dimensions comes from the internal Z boson
KK modes after the compactification of the external dimensions on
orbifold $S^1/Z_2$ (($S^1\times S^1)/Z_2$) for a single (two)
extra dimensions. We observe that there is almost two orders
larger enhancement on the BR of the process we study in the case
of two NUED. This enhancement is obviously due to the abundance of
$Z$  boson KK modes.

The paper is organized as follows: In Section 2, we present the
theoretical expression for the decay width of the LFV decay
$\tau\rightarrow \mu \,\bar{\nu_i}\, \nu_i$, $i=e,\mu,\tau$, in
the framework of the model III, with the inclusion of one and two
NUEDs. Section 3 is devoted to discussion and our conclusions. In
the appendix section, we give explicit expressions of the
functions appearing in the general effective vertex for the
interaction of off-shell Z-boson with a fermionic current.
\section{$\tau\rightarrow \mu \, \bar{\nu_i} \, \nu_i$ decay in the general
two Higgs doublet model with the inclusion of non-universal extra
dimensions}
The LFV  $\tau\rightarrow \mu \, \bar{\nu_i} \, \nu_i$ decay
exists at least in the one loop level and, therefore, the physical
quantities like the BR contains rich information about the model
used and the free parameters existing. The extension of the Higgs
sector in the SM makes the flavor violation (FV) possible with the
help of the new Yukawa interactions coming from the new Higgs
scalars. In the multi Higgs doublet models, the flavor changing
neutral current (FCNC) at tree level, which induces the FV
interactions, can appear and the general 2HDM, so called model
III, is one of the candidate. The additional Yukawa interactions
arising from new Higgs doublet are responsible for the non
vanishing theoretical values of the physical quantities like the
BR.  The inclusion of extra dimensions may bring further
contributions to these physical quantities. In the present work we
assume that the first Higgs doublet and the gauge fields feel the
extra dimensions, however, the leptons, the quarks and the second
Higgs doublet, which contains new Higgs particles, live in the 4D
brane. With the addition of single (double) extra dimension, the
Yukawa interaction, including the LFV part in the model III, reads
\begin{eqnarray}
{\cal{L}}_{Y}= \eta^{E}_{5 (6)\, ij} \bar{l}_{i L} (\phi_{1}|_{y
(z)=0}) E_{j R}+ \xi^{E}_{ij} \bar{l}_{i L} \phi_{2} E_{j R} +
h.c. \,\,\, , \label{lagrangian}
\end{eqnarray}
where $i,j$ are family indices of leptons, $L$ and $R$ denote
chiral projections $L(R)=1/2(1\mp \gamma_5)$, $\phi_{i}$ for
$i=1,2$, are the two scalar doublets, $l_{i L}$ and $E_{j R}$ are
lepton doublets and singlets respectively. Here $\eta^{E}_{5 (6)\,
ij}$ are $5 (6)$-dimensional dimensionful, $\xi^{E}_{ij}$ are
dimensionless Yukawa couplings and  $\eta^{E}_{5 (6)\, ij}$ can be
rescaled to the ones in 4-dimension as $\eta^{E}_{5 (6)\,
ij}=\sqrt{2 \pi R}\,(2 \pi R)\, \eta^{E}_{ij}$, $R$ is the
compactification radius and $y$ ($z$) is the coordinate represents
the $5 (6)$'th dimension. The coupling $\xi^{E}_{ij}$, which has
complex entries in general, is responsible for the FCNC at tree
level \footnote{ Notice that, in the following, we replace
$\xi^{E}$ with $\xi^{E}_{N}$ where "N" denotes the word
"neutral".}. At this stage we take two Higgs doublets $\phi_{1}$
and $\phi_{2}$ as
\begin{eqnarray}
\phi_{1}=\frac{1}{\sqrt{2}}\left[\left(\begin{array}{c c}
0\\v+H^{0}\end{array}\right)\; + \left(\begin{array}{c c}
\sqrt{2} \chi^{+}\\ i \chi^{0}\end{array}\right) \right]\, ;
\phi_{2}=\frac{1}{\sqrt{2}}\left(\begin{array}{c c}
\sqrt{2} H^{+}\\ H_1+i H_2 \end{array}\right) \,\, ,
\label{choice}
\end{eqnarray}
respecting that only the first one has a vacuum expectation value
but not the second, namely,
\begin{eqnarray}
<\phi_{1}>=\frac{1}{\sqrt{2}}\left(\begin{array}{c c}
0\\v\end{array}\right) \,  \, ;
<\phi_{2}>=0 \,\, .
\label{choice2}
\end{eqnarray}
%
%
%
Therefore the CP even neutral Higgs particles $H_1$ and $H^0$ do
not mix in the tree level and they are obtained as the mass
eigenstates $h^0$ and $H^0$ respectively. The neutral Higgs
particle $H_2$ is the well known CP odd $A^0$. In this case, the
SM particles lie in the first doublet and the new particles in the
second one.

\begin{figure}[htb]
\vskip -0.5truein \centering \epsfxsize=6.0in
\leavevmode\epsffile{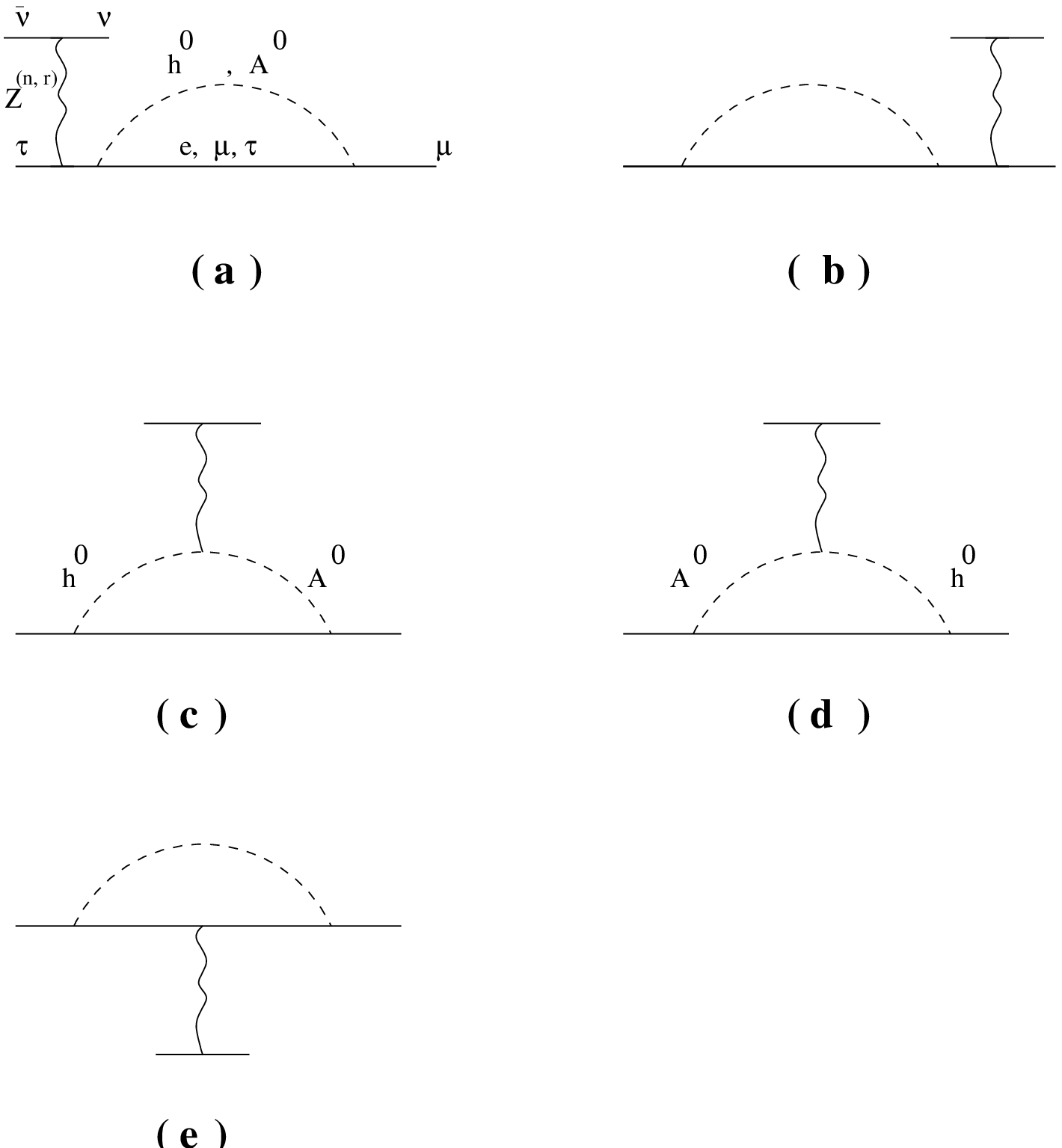} \vskip 0.5truein \caption[]{One
loop diagrams contribute to $\tau\rightarrow \mu \,\bar{\nu_i}\,
\nu_i$, $i=e,\mu,\tau$ decay due to the neutral Higgs bosons $h_0$
and $A_0$ in the model III version of 2HDM. Solid lines represent
leptons and neutrinos, curly (dashed) lines represent the virtual
$Z$ boson and its KK modes in 6 dimensions ($h_0$ and $A_0$
fields).} \label{fig1ver}
\end{figure}

The LFV  $\tau\rightarrow \mu \, \bar{\nu_i} \, \nu_i$ decay is
induced by $\tau\rightarrow \mu  Z^*$ transition and  $Z^* \,
\rightarrow \bar{\nu_i} \, \nu_i$ process (see Fig.
\ref{fig1ver}). The $\tau\rightarrow \mu Z^*$ transition, which
needs the FCNC at tree level, is driven by the internal new
neutral Higgs bosons $h^0$ and $A^0$, which are living in the 4D
brane. Since the gauge fields are accessible to the extra
dimensions, the internal $Z$ boson has KK modes, which have
additional contribution to the physical quantities related to the
decay studied. The KK modes of gauge fields appear after the
compactification of the external dimensions on the orbifold
$S^1/Z_2$ ( ($S^1\times S^1)/Z_2$ ) for a single (two) extra
dimensions and, for two extra dimensions, the gauge fields can be
expanded to the KK modes as:
\begin{eqnarray}
A_\mu (x,y,z ) & = & \frac{1}{(2 \pi R)^{d/2}} \left\{
A_\mu^{(0,0)}(x) + 2^{d/2} \sum_{n,r}^{\infty} A_\mu^{(n,r)}(x)
\cos(ny/R+rz/R)\right\} \, , \nonumber \\
A_{i} (x,y,z ) & = & \frac{1}{(2 \pi R)^{d/2}} \left\{ 2^{d/2}
\sum_{n,r}^{\infty} A_i^{(n,r)}(x) \sin (ny/R+rz/R)\right\} \, ,
\label{Gauge1}
\end{eqnarray}
where $i=5,6$, $d=2$, the indices  $n$ and $r$ are positive
integers including zero, but both are not zero at the same time.
The KK modes of the gauge field $Z$  have the masses
$\sqrt{m_{Z}^2+m_n^2+m_r^2}$, where $m_n=n/R$ and $m_r=r/R$.
Notice that the gauge boson mass matrix is diagonal since the
Higgs field, which has the non-zero vacuum expectation value, is
in the bulk (see \cite{Ruckl} for details). Here, we take the
compactification radius $R$ as the same for both new dimensions.
In the case of a single extra dimension, one should set $d=1$,
take $z=0$, and drop the summation over $r$ in eq. (\ref{Gauge1}).

Now, we present the general effective vertex for the interaction
of off-shell Z-boson with a fermionic current
\begin{equation}
\Gamma^{(REN)}_\mu (\tau\rightarrow \mu Z^*)=  f_1 \, \gamma_\mu +
f_2  \,\gamma_\mu \gamma_5+f_3\, \sigma_{\mu\nu}
k^\nu+f_4\,\sigma_{\mu\nu}\gamma_5 k^\nu  \, ,\label{GammaRen2}
\end{equation}
where $k$ is the momentum transfer, $k^2=(p-p')^2$, $p$
($p^{\prime}$) is the four momentum vector of incoming (outgoing)
lepton and the explicit expressions for the functions $f_1$,
$f_2$, $f_3$ and $f_4$ are given in the appendix section. The
matrix element $M$ of the $\tau\rightarrow \mu \, \bar{\nu_i}\,
\nu_i$ process is obtained with the internal $Z$ boson connection
between the $\tau\rightarrow \mu $ transition and the $\bar{\nu_i}
\nu_i$ pair. Since the gauge fields are accessible to the extra
dimensions, the KK modes of the internal $Z$ boson have additional
contributions to the process (see Fig. \ref{fig1ver}). Notice that
the KK number need not to be conserved in the vertices where $Z$
boson appears since the the extra dimensions in this case are so
called NUED. Using the matrix element $M$ the decay width $\Gamma$
of the decay under consideration can be obtained in the $\tau$
lepton rest frame with the help of the well known expression
\begin{equation}
d\Gamma=\frac{(2\, \pi)^4}{2\, m_\tau} \, |M|^2\,\delta^4
(p-\sum_{i=1}^3 p_i)\,\prod_{i=1}^3\,\frac{d^3p_i}{(2 \pi)^3 2
E_i} \,
 ,
\label{DecWidth}
\end{equation}
where $p$ ($p_i$, i=1,2,3) is four momentum vector of $\tau$
lepton ($\mu$ lepton, incoming $\nu$, outgoing $\nu$).
\section{Discussion}
The $\tau\rightarrow \mu \, \bar{\nu_i} \, \nu_i$ is induced by
the LFV $\tau\rightarrow \mu$ transition which depends on the
various Yukawa couplings $\bar{\xi}^E_{N,ij}$
\footnote{The dimensionfull Yukawa couplings
$\bar{\xi}^{E}_{N,ij}$  are defined as
$\xi^{E}_{N,ij}=\sqrt{\frac{4\,G_F}{\sqrt{2}}}\,
\bar{\xi}^{E}_{N,ij}$.}
$i=\mu,\tau$; $j=e,\mu,\tau$  and they need to be restricted by
using the present and forthcoming experiments. In our work, we
take only the $\tau$ lepton as an internal lepton and, therefore,
we choose the couplings $\bar{\xi}_{N,\tau\tau}^E$ and
$\bar{\xi}_{N,\tau\mu}^E$ as non-zero. Here we expect that the
couplings which contain at least one $\tau$ index are dominant
similar to the Cheng-Sher scenario \cite{Sher}. The upper limit of
the coupling $\bar{\xi}^{E}_{N,\tau \mu}$ has been estimated as
$30\, GeV$ (see \cite{Iltananomuon} and references therein) by
assuming that the new physics effects can not exceed experimental
uncertainty $10^{-9}$ in the measurement of the muon anomalous
magnetic moment. In our numerical calculation we choose
$\bar{\xi}^{E}_{N,\tau \mu}=1\,GeV$ by respecting this upper
limit. Since there is no restriction for the Yukawa coupling
$\bar{\xi}^{E}_{N,\tau \tau}$, the numerical values we use are
greater than $\bar{\xi}^{E}_{N,\tau \mu}$.

This work is devoted to the a single and two NUED effects on the
BR of the LFV processes $\tau\rightarrow \mu \, \bar{\nu_i} \,
\nu_i$, in the type III 2HDM. In the case of two extra dimensions,
we observe that the contribution of KK modes enhances the BR
considerably, due to the crowd of Z boson KK modes. Notice that we
use the numerical values, $m_Z= 91\, (GeV)$, $m_W= 80\, (GeV)$,
$s_w=\sqrt 0.23$, $G_F=1.6637 \times 10^{-5}\, (GeV^{-2})$,
$\Gamma_\tau=2.27 \times 10^{-12}\, (GeV)$, in our numerical
calculations.

\begin{figure}[htb]
\vskip -3.0truein \centering \epsfxsize=6.8in
\leavevmode\epsffile{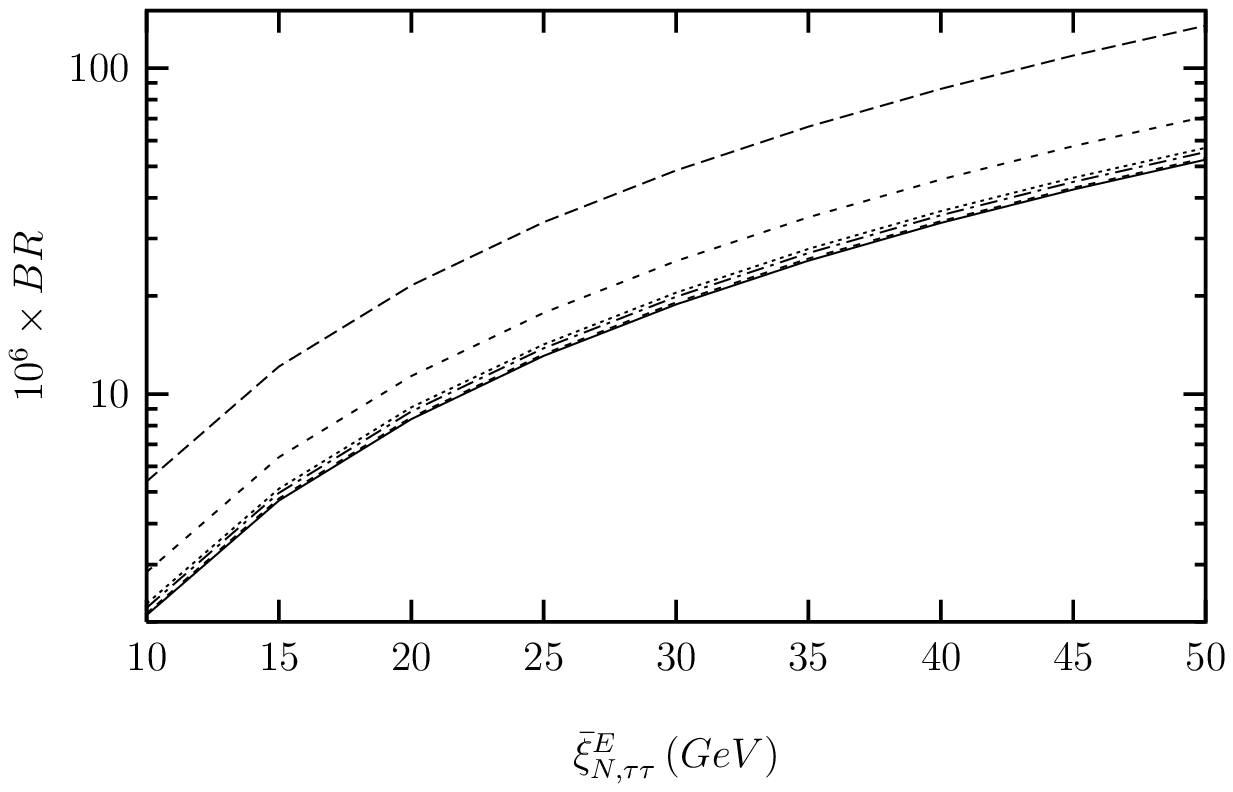} \vskip -3.0truein \caption[]{
$\bar{\xi}^{E}_{N,\tau \tau}$ dependence of the BR for
$\bar{\xi}^{E}_{N,\tau \mu}=1\, (GeV)$ and for different values of
the compactification scale $1/R$, in the case of a single extra
dimension. Here solid (dashed, small dashed, dotted, dash-dotted,
three dashed-spaced) line represents the BR without extra
dimensions (with extra dimensions for $1/R=200, 400, 800, 1000,
2000\, (GeV)$). } \label{BR01tautau}
\end{figure}

In Fig. \ref{BR01tautau}, we present $\bar{\xi}^{E}_{N,\tau \tau}$
dependence of the BR for $\bar{\xi}^{E}_{N,\tau \mu}=1\, (GeV)$
and for different values of the compactification scale $1/R$, in
the case of a single extra dimension. Here solid (dashed, small
dashed, dotted, dash-dotted, three dashed-spaced) line represents
the case without extra dimensions (with extra dimensions for
$1/R=200, 400, 800, 1000, 2000\, (GeV)$). It is observed that the
BR lies in the range $1.0\times 10^{-6}-5.0\times 10^{-5}$ for the
interval of the Yukawa coupling $10\, (GeV)\leq
\bar{\xi}^{E}_{N,\tau \tau} \leq 50\, (GeV)$. The addition of a
single extra dimension enhances the $BR$ almost twice for the
small values of the compactification scale. However this
enhancement is not more than the order of $1\%$ for its large
values.

\begin{figure}[htb]
\vskip -3.0truein \centering \epsfxsize=6.8in
\leavevmode\epsffile{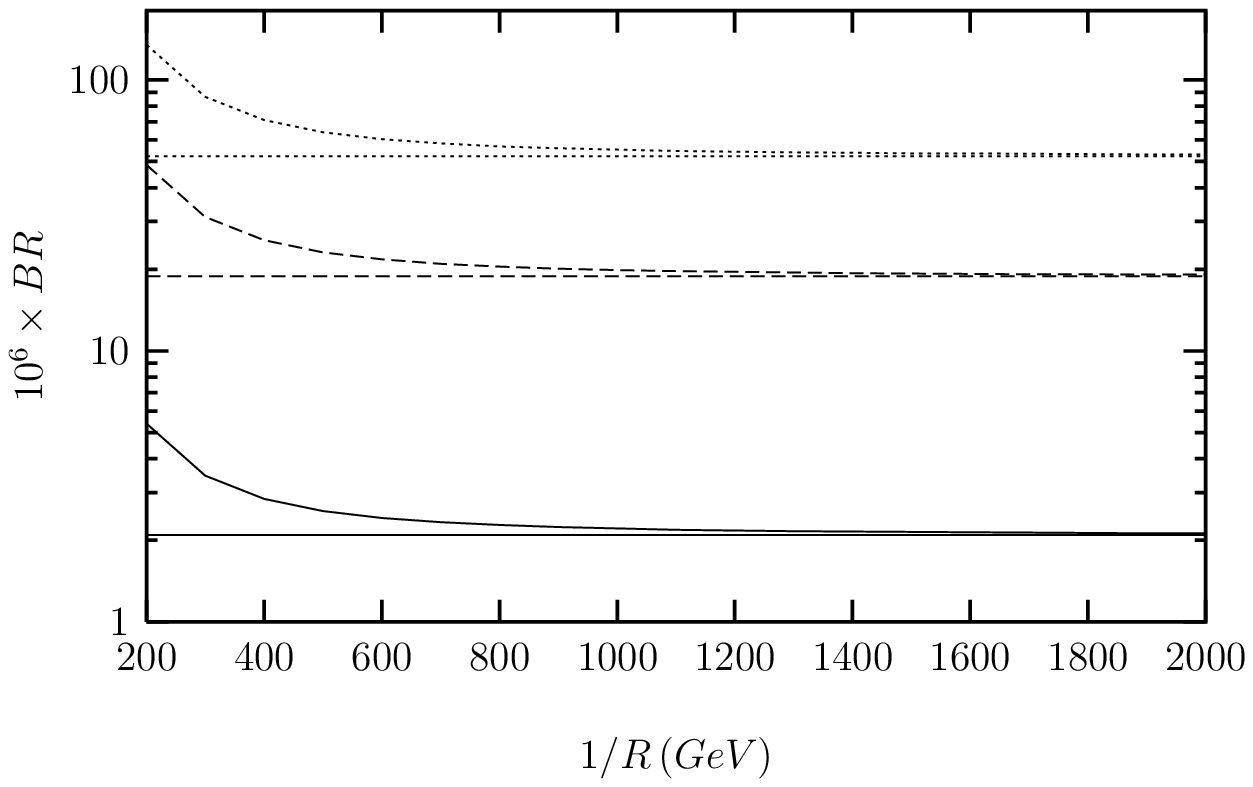} \vskip -3.0truein \caption[]{The
compactification scale $1/R$ dependence of the BR for
$\bar{\xi}^{E}_{N,\tau \mu}=1\, (GeV)$ and for different values of
the coupling $\bar{\xi}^{E}_{N,\tau \tau}$, in the case of a
single extra dimension. Here solid-dashed-dotted, straight lines
(curved lines) represents the BR for $\bar{\xi}^{E}_{N,\tau
\tau}=10-30-50\, (GeV)$ without extra dimensions (with extra
dimensions).} \label{BR01Rr}
\end{figure}

Fig. \ref{BR01Rr}, represents the compactification scale $1/R$
dependence of the BR for $\bar{\xi}^{E}_{N,\tau \mu}=1\, (GeV)$
and for different values of the coupling $\bar{\xi}^{E}_{N,\tau
\tau}$, in the case of a single extra dimension. Here
solid-dashed-dotted straight lines (curved lines) represents the
BR for $\bar{\xi}^{E}_{N,\tau \tau}=10-30-50\, (GeV)$ without
extra dimensions (with extra dimensions). This figure also shows
that the contribution due to the extra dimensions is negligible
especially for $1/R \geq 600 \, (GeV)$.

Now we make the same analysis for two NUED. In this case the crowd
of Z boson KK modes cause to enhance the BR of the decay analyzed
and these effects can be observable. Notice that there is a
possible divergence problem due to the abundance of KK modes in
the summation of the KK mode contributions, however, the ratio
$\frac{1}{m_{Z}^2+(n^2+r^2)/R^2}$ appearing in the internal line
converges to zero sharply with the increasing values of the
integers $n$ and $r$ and the convergence of the KK sum is obtained
for the region of the compactification scale that we study, $1/R >
200\, GeV$.

\begin{figure}[htb]
\vskip -3.0truein \centering \epsfxsize=6.8in
\leavevmode\epsffile{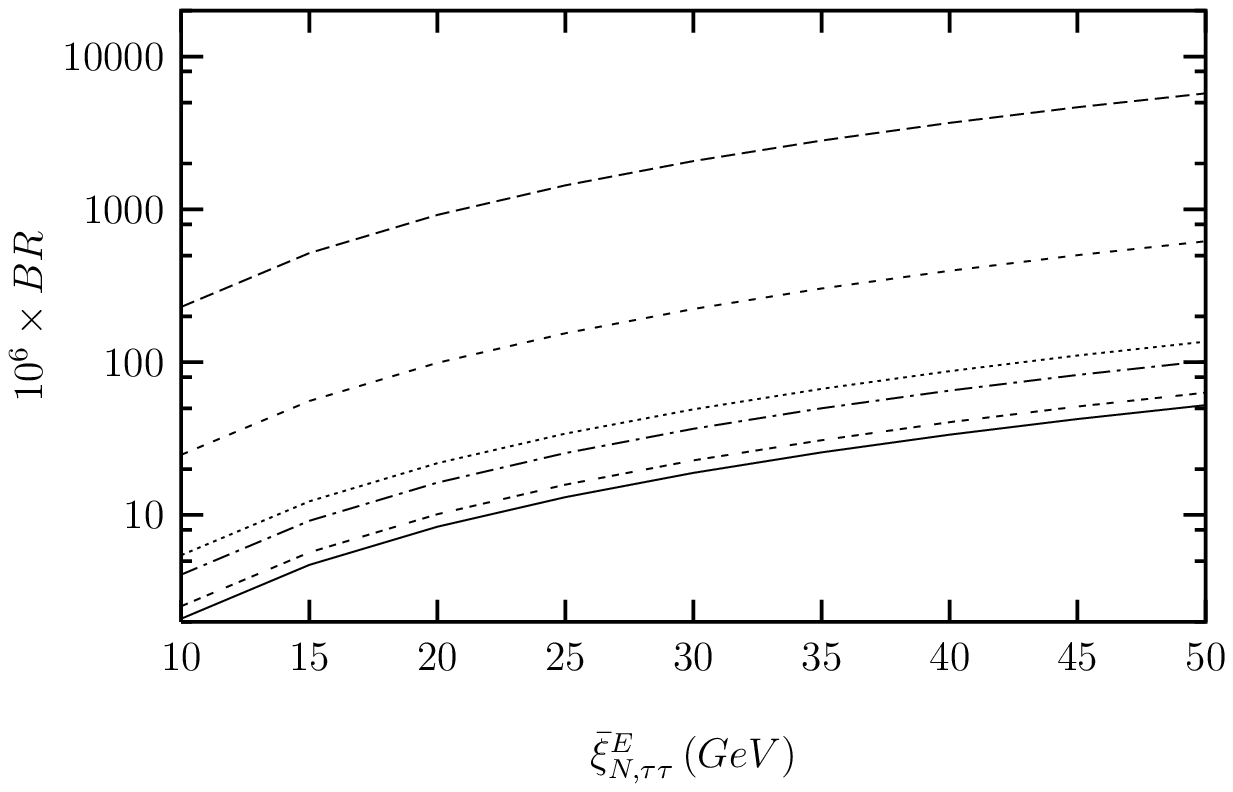} \vskip -3.0truein
\caption[]{$\bar{\xi}^{E}_{N,\tau \tau}$ dependence of the BR for
$\bar{\xi}^{E}_{N,\tau \mu}=1\, (GeV)$ and for different values of
the compactification scale $1/R$, in the case of two extra
dimensions. Here solid (dashed, small dashed, dotted, dash-dotted,
three dashed-spaced) line represents the BR without extra
dimensions (with extra dimensions for $1/R=200, 400, 800, 1000,
2000\, GeV$) } \label{BR02tautau}
\end{figure}

In Fig. \ref{BR02tautau}, we present $\bar{\xi}^{E}_{N,\tau \tau}$
dependence of the BR for $\bar{\xi}^{E}_{N,\tau \mu}=1\, (GeV)$
and for different values of the compactification scale $1/R$ in
the case of two NUEDs. Here solid (dashed, small dashed, dotted,
dash-dotted, three dashed-spaced) line represents the case without
extra dimensions (with extra dimensions for $1/R=200, 400, 800,
1000, 2000\, (GeV)$). This figure shows that the BR is
considerably enhanced for the small values of the compactification
scale $1/R$. The BR is two orders (one order, three times, two
times, 30 \%) larger for $1/R=200\, (GeV)$ ($1/R=400\, (GeV)$,
$1/R=800\, (GeV)$, $1/R=1000\, (GeV)$, $1/R=2000\, (GeV)$)
compared to the one without NUED. Even for large values of the
compactification scale near $1/R\sim 2000\, (GeV)$ there is an
enhancement in the BR.

\begin{figure}[htb]
\vskip -3.0truein \centering \epsfxsize=6.8in
\leavevmode\epsffile{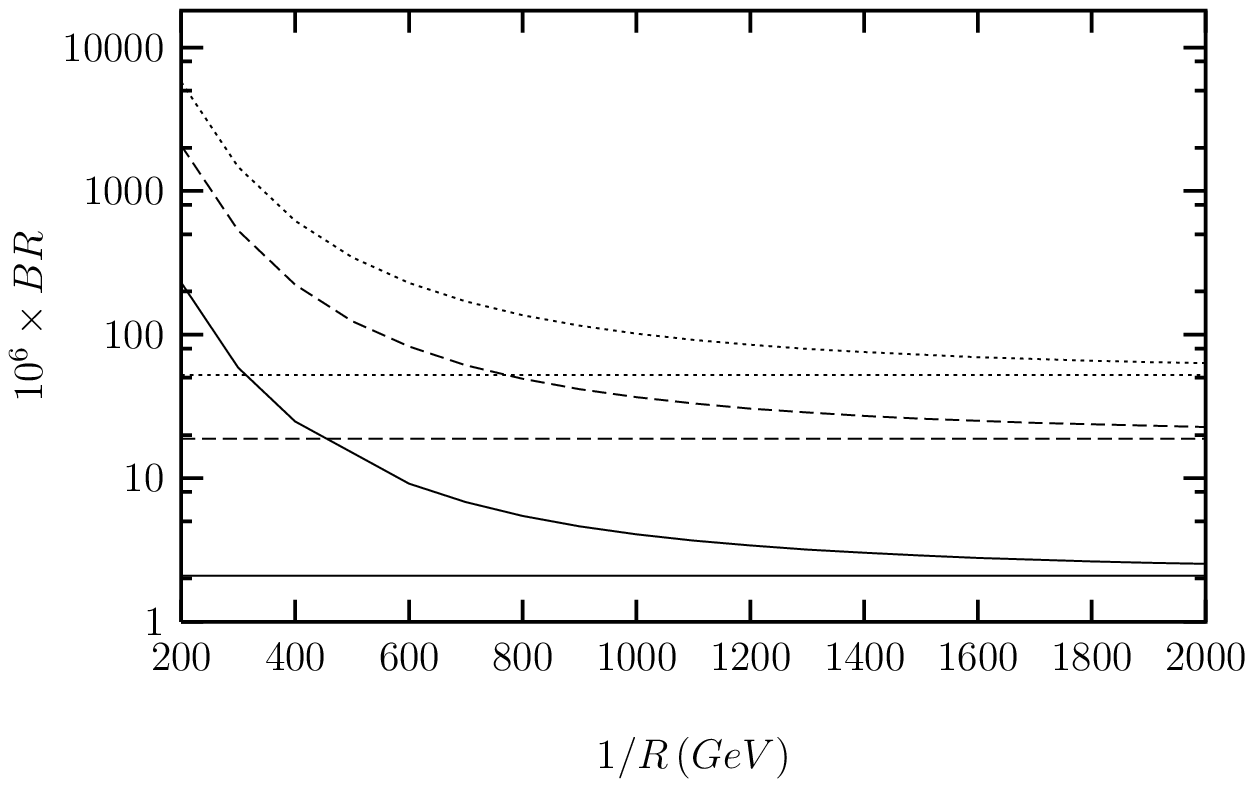} \vskip -3.0truein \caption[]{The
compactification scale $1/R$ dependence of the $BR$ for
$\bar{\xi}^{E}_{N,\tau \mu}=1\, (GeV)$ and for different values of
the coupling $\bar{\xi}^{E}_{N,\tau \tau}$, in the case of two
extra dimensions. Here solid-dashed-dotted, straight lines (curved
lines) represents the BR for $\bar{\xi}^{E}_{N,\tau
\tau}=10-30-50\, (GeV)$ without extra dimensions (with extra
dimensions).} \label{BR02Rr}
\end{figure}

Fig. \ref{BR02Rr} is devoted to the compactification scale $1/R$
dependence of the $BR$ for $\bar{\xi}^{E}_{N,\tau \mu}=1\, (GeV)$
and for different values of the coupling $\bar{\xi}^{E}_{N,\tau
\tau}$, for two NUEDs. Here solid-dashed-dotted straight lines
(curved lines) represents the BR for $\bar{\xi}^{E}_{N,\tau
\tau}=10-30-50\, (GeV)$ without extra dimensions (with extra
dimensions). The enhancement in the BR is observed in this figure
also. These contributions become negligible for the large values
of the compactification scale, namely for $1/R > 2000 \, (GeV)$.

At this stage we would like to summarize our results:
\begin{itemize}
\item  We predict the BR in the range $1.0\times10^{-6}-5.0\times
10^{-5}$ for the interval of the Yukawa coupling $10\, (GeV)\leq
\bar{\xi}^{E}_{N,\tau \tau} \leq 50\, (GeV)$. The inclusion of a
single NUED brings contributions which increase the BR twice of
the one without the extra dimension, for the small values of the
compactification scale. However this enhancement  is not more than
the order of $1\%$ for the large values of the compactification
scale.
\item  With the inclusion of two extra dimensions  the BR is
considerably enhanced for the small values of the compactification
scale $1/R$, even two orders. In the case of large values of the
compactification scale, there is still an enhancement in the BR.
\end{itemize}

Therefore, the future theoretical and experimental investigations
of the process $\tau\rightarrow \mu  \, \bar{\nu}_i \nu_i$ would
ensure a valuable information about signals coming from the extra
dimensions.
\newpage
\section{Appendix}
The explicit expressions for the functions $f_1$, $f_2$, $f_3$ and
$f_4$ appearing in eq. (\ref{GammaRen2}) read
\begin{eqnarray}
f_1&=& \frac{g}{64\,\pi^2\,cos\,\theta_W} \int_0^1\, dx \,
\frac{1}{m^2_{l_2}-m^2_{l_1}} \Bigg \{ c_V \, (m_{l_2}+m_{l_1})
\nonumber \\
&\Bigg(& (-m_i \, \eta^+_i + m_{l_1} (-1+x)\, \eta_i^V)\, ln \,
\frac{L^{self}_ {1,\,h^0}}{\mu^2}+ (m_i \, \eta^+_i - m_{l_2}
(-1+x)\, \eta_i^V)\, ln \, \frac{L^{self}_{2,\, h^0}}{\mu^2}
\nonumber \\ &+& (m_i \, \eta^+_i + m_{l_1} (-1+x)\, \eta_i^V)\,
ln \, \frac{L^{self}_{1,\, A^0}}{\mu^2} - (m_i \, \eta^+_i +
m_{l_2} (-1+x) \,\eta_i^V)\, ln \, \frac{L^{self}_{2,\,
A^0}}{\mu^2} \Bigg) \nonumber \\ &+&
c_A \, (m_{l_2}-m_{l_1}) \nonumber \\
&\Bigg ( & (-m_i \, \eta^-_i + m_{l_1} (-1+x)\, \eta_i^A)\, ln \,
\frac{L^{self}_{1,\, h^0}}{\mu^2} + (m_i \, \eta^-_i + m_{l_2}
(-1+x)\, \eta_i^A)\, ln \, \frac{L^{self}_{2,\, h^0}}{\mu^2}
\nonumber \\ &+& (m_i \, \eta^-_i + m_{l_1} (-1+x)\, \eta_i^A)\,
ln \, \frac{L^{self}_{1,\, A^0}}{\mu^2} + (-m_i \, \eta^-_i +
m_{l_2} (-1+x)\, \eta_i^A)\, ln \, \frac{L^{self}_{2,\,
A^0}}{\mu^2} \Bigg) \Bigg \} \nonumber \\ &-&
\frac{g}{64\,\pi^2\,cos\,\theta_W} \int_0^1\,dx\, \int_0^{1-x} \,
dy \, \Bigg \{ m_i^2 \,(c_A\,
\eta_i^A-c_V\,\eta_i^V)\,(\frac{1}{L^{ver}_{A^0}}+
\frac{1}{L^{ver}_{h^0}}) \nonumber \\ &-& (1-x-y)\,m_i\, \Bigg(
c_A\,  (m_{l_2}-m_{l_1})\, \eta_i^- \,(\frac{1}{L^{ver}_{h^0}} -
\frac{1}{L^{ver}_{A^0}})+ c_V\, (m_{l_2}+m_{l_1})\, \eta_i^+ \,
(\frac{1}{L^{ver}_{h^0}} + \frac{1}{L^{ver}_{A^0}}) \Bigg)
\nonumber \\ &-& (c_A\, \eta_i^A+c_V\,\eta_i^V) \Bigg (
-2+(k^2\,x\,y+m_{l_1}\,m_{l_2}\, (-1+x+y)^2)\,
(\frac{1}{L^{ver}_{h^0}} +
\frac{1}{L^{ver}_{A^0}})-ln\,\frac{L^{ver}_{h^0}}{\mu^2}\,
\frac{L^{ver}_{A^0}}{\mu^2} \Bigg ) \nonumber \\ &-&
(m_{l_2}+m_{l_1})\, (1-x-y)\, \Bigg ( \frac{\eta_i^A\,(x\,m_{l_1}
+y\,m_{l_2})+m_i\,\eta_i^-}
{2\,L^{ver}_{A^0\,h^0}}+\frac{\eta_i^A\,(x\,m_{l_1} +y\,m_{l_2})-
m_i\,\eta_i^-}{2\,L^{ver}_{h^0\,A^0}} \Bigg ) \nonumber \\ &+&
\frac{1}{2}\eta_i^A\, ln\,\frac{L^{ver}_{A^0\,h^0}}{\mu^2}\,
\frac{L^{ver}_{h^0\,A^0}}{\mu^2}
\Bigg \}\,, \nonumber \\
f_2&=& \frac{g}{64\,\pi^2\,cos\,\theta_W} \int_0^1\, dx \,
\frac{1}{m^2_{l_2}-m^2_{l_1}} \Bigg \{ c_V \, (m_{l_2}-m_{l_1})
\nonumber \\
&\Bigg(& (m_i \, \eta^-_i + m_{l_1} (-1+x)\, \eta_i^A)\, ln \,
\frac{L^{self}_{1,\,A^0}}{\mu^2} + (-m_i \, \eta^-_i + m_{l_2}
(-1+x)\, \eta_i^A)\, ln \, \frac{L^{self}_ {2,\,A^0}}{\mu^2}
\nonumber \\ &+& (-m_i \, \eta^-_i + m_{l_1} (-1+x)\, \eta_i^A)\,
ln \, \frac{L^{self}_{1,\, h^0}}{\mu^2}+ (m_i \, \eta^-_i +
m_{l_2} (-1+x)\, \eta_i^A)\, ln \,
\frac{L^{self}_{2,\,h^0}}{\mu^2} \Bigg) \nonumber \\ &+&
c_A \, (m_{l_2}+m_{l_1}) \nonumber \\
&\Bigg(& (m_i \, \eta^+_i + m_{l_1} (-1+x)\, \eta_i^V)\, ln \,
\frac{L^{self}_{1,\, A^0}}{\mu^2}- (m_i \, \eta^+_i + m_{l_2}
(-1+x)\, \eta_i^V)\, ln \, \frac{L^{self}_{2,\,A^0}}{\mu^2}
\nonumber \\ &+& (-m_i \, \eta^+_i + m_{l_1} (-1+x)\, \eta_i^V)\,
ln \, \frac{L^{self}_{1,\, h^0}}{\mu^2} + (m_i \, \eta^+_i -
m_{l_2} (-1+x)\, \eta_i^V)\, \frac{ln \,
L^{self}_{2,\,h^0}}{\mu^2} \Bigg) \Bigg \} \nonumber \\ &+&
\frac{g}{64\,\pi^2\,cos\,\theta_W} \int_0^1\,dx\, \int_0^{1-x} \,
dy \, \Bigg \{ m_i^2 \,(c_V\,
\eta_i^A-c_A\,\eta_i^V)\,(\frac{1}{L^{ver}_{A^0}}+
\frac{1}{L^{ver}_{h^0}}) \nonumber \\ &-& m_i\, (1-x-y)\, \Bigg(
c_V\, (m_{l_2}-m_{l_1}) \,\eta_i^- + c_A\, (m_{l_2}+m_{l_1})\,
\eta_i^+ \Bigg) \,(\frac{1} {L^{ver}_{h^0}} -
\frac{1}{L^{ver}_{A^0}}) \nonumber \\ &+& (c_V\,
\eta_i^A+c_A\,\eta_i^V) \Bigg(-2+(k^2\,x\,y-m_{l_1}\,m_{l_2}\,
(-1+x+y)^2) (\frac{1}{L^{ver}_{h^0}}+\frac{1}{L^{ver}_{A^0}})-
ln\,\frac{L^{ver}_{h^0}}{\mu^2}\,\frac{L^{ver}_{A^0}}{\mu^2}
\Bigg) \nonumber \\ &-& (m_{l_2}-m_{l_1})\, (1-x-y)\, \Bigg(
\frac{\eta_i^V\,(x\,m_{l_1} -y\,m_{l_2})+m_i\,\eta_i^+}
{2\,L^{ver}_{A^0\,h^0}}+ \frac{\eta_i^V\,(x\,m_{l_1}
-y\,m_{l_2})-m_i\, \eta_i^+}{2\,L^{ver}_{h^0\,A^0}}
\Bigg)\nonumber \\
&-& \frac{1}{2} \eta_i^V\, ln\,\frac{L^{ver}_{A^0\,h^0}}{\mu^2}\,
\frac{L^{ver}_{h^0\,A^0}}{\mu^2}
\Bigg \} \nonumber \,,\\
f_3&=&-i \frac{g}{64\,\pi^2\,cos\,\theta_W} \int_0^1\,dx\,
\int_0^{1-x} \, dy \, \Bigg \{ \Bigg( (1-x-y)\,(c_V\,
\eta_i^V+c_A\,\eta_i^A)\, (x\,m_{l_1} +y\,m_{l_2}) \nonumber
\\ &+& \, m_i\,(c_A\, (x-y)\,\eta_i^-+c_V\,\eta_i^+\,(x+y))\Bigg )
\,\frac{1}{L^{ver}_{h^0}} \nonumber \\ &+& \Bigg( (1-x-y)\, (c_V\,
\eta_i^V+c_A\,\eta_i^A)\, (x\,m_{l_1} +y\,m_{l_2}) -m_i\,(c_A\,
(x-y)\,\eta_i^-+c_V\,\eta_i^+\,(x+y))\Bigg )
\,\frac{1}{L^{ver}_{A^0}} \nonumber \\ &-& (1-x-y) \Bigg
(\frac{\eta_i^A\,(x\,m_{l_1} +y\,m_{l_2})}{2}\, \Big (
\frac{1}{L^{ver}_{A^0\,h^0}}+\frac{1}{L^{ver}_{h^0\,A^0}} \Big )
+\frac{m_i\,\eta_i^-} {2} \, \Big ( \frac{1}{L^{ver}_{h^0\,A^0}}-
\frac{1}{L^{ver}_{A^0\,h^0}} \Big ) \Bigg ) \Bigg \} \,,\nonumber \\
f_4&=&-i\frac{g}{64\,\pi^2\, cos\,\theta_W} \int_0^1\,dx\,
\int_0^{1-x} \, dy \, \Bigg \{ \Bigg( (1-x-y)\,\Big ( -(c_V\,
\eta_i^A+c_A\,\eta_i^V)\, (x\,m_{l_1} -y\, m_{l_2}) \Big)
\nonumber \\ &-& m_i\, (c_A\,
(x-y)\,\eta_i^++c_V\,\eta_i^-\,(x+y))\Bigg )\,
\frac{1}{L^{ver}_{h^0}} \nonumber \\ &+& \Bigg ( (1-x-y)\,\Big (
-(c_V\, \eta_i^A+c_A\,\eta_i^V)\, (x\,m_{l_1} - y\, m_{l_2}) \Big
) + m_i\,(c_A\, (x-y)\,\eta_i^++c_V\,\eta_i^-\,(x+y)) \Bigg )
\,\frac{1}{L^{ver}_{A^0}} \nonumber \\&+& (1-x-y)\, \Bigg (
\frac{\eta_i^V}{2}\,(m_{l_1}\,x-m_{l_2}\, y)\, \, \Big (
\frac{1}{L^{ver}_{A^0\,h^0}}+\frac{1}{L^{ver}_{h^0\,A^0}} \Big )
+\frac{m_i\,\eta_i^+}{2}\, \Big (
\frac{1}{L^{ver}_{A^0\,h^0}}-\frac{1}{L^{ver}_{h^0\,A^0}} \Big )
\Bigg ) \Bigg \}\, , \label{fVAME}
\end{eqnarray}
where
\begin{eqnarray}
L^{self}_{1,\,h^0}&=&m_{h^0}^2\,(1-x)+(m_i^2-m^2_{l_1}\,(1-x))\,x
\nonumber \, , \\
L^{self}_{1,\,A^0}&=&L^{self}_{1,\,h^0}(m_{h^0}\rightarrow
m_{A^0})
\nonumber \, , \\
L^{self}_{2,\,h^0}&=&L^{self}_{1,\,h^0}(m_{l_1}\rightarrow
m_{l_2})
\nonumber \, , \\
L^{self}_{2,\,A^0}&=&L^{self}_{1,\,A^0}(m_{l_1}\rightarrow
m_{l_2})
\nonumber \, , \\
L^{ver}_{h^0}&=&m_{h^0}^2\,(1-x-y)+m_i^2\,(x+y)-k^2\,x\,y
\nonumber \, , \\
L^{ver}_{h^0\,A^0}&=&m_{A^0}^2\,x+m_i^2\,(1-x-y)+(m_{h^0}^2-k^2\,x)\,y
\nonumber \, , \\
L^{ver}_{A^0}&=&L^{ver}_{h^0}(m_{h^0}\rightarrow m_{A^0})
\nonumber \, , \\
L^{ver}_{A^0\,h^0}&=&L^{ver}_{h^0\,A^0}(m_{h^0}\rightarrow
m_{A^0}) \, , \label{Lh0A0}
\end{eqnarray}
and
\begin{eqnarray}
\eta_i^V&=&\xi^{E}_{N,l_1i}\xi^{E\,*}_{N,il_2}+
\xi^{E\,*}_{N,il_1} \xi^{E}_{N,l_2 i} \nonumber \, , \\
\eta_i^A&=&\xi^{E}_{N,l_1i}\xi^{E\,*}_{N,il_2}-
\xi^{E\,*}_{N,il_1} \xi^{E}_{N,l_2 i} \nonumber \, , \\
\eta_i^+&=&\xi^{E\,*}_{N,il_1}\xi^{E\,*}_{N,il_2}+
\xi^{E}_{N,l_1i} \xi^{E}_{N,l_2 i} \nonumber \, , \\
\eta_i^-&=&\xi^{E\,*}_{N,il_1}\xi^{E\,*}_{N,il_2}-
\xi^{E}_{N,l_1i} \xi^{E}_{N,l_2 i}\, . \label{etaVA}
\end{eqnarray}
The parameters $c_V$ and $c_A$ are $c_A=-\frac{1}{4}$ and
$c_V=\frac{1}{4}-sin^2\,\theta_W$. In eq. (\ref{etaVA}) the flavor
changing couplings $\xi^{E}_{N, ji}$ represent the effective
interaction between the internal lepton $i$, ($i=e,\mu,\tau$) and
outgoing (incoming) $j=l_1\,(j=l_2)$ one. Here we take only the
$\tau$ lepton in the internal line and  we neglect all the Yukawa
couplings except $\xi_{N,\tau\tau}^E$ and $\xi_{N,\tau\mu}^E$ in
the loop contributions (see Discussion section). The  Yukawa
couplings $\xi^{E}_{N, ji}$ are complex in general, however, in
the present work, we take them real.
\section{Acknowledgement}
This work has been supported by the Turkish Academy of Sciences in
the framework of the Young Scientist Award Program.
(EOI-TUBA-GEBIP/2001-1-8)

\newpage
\newpage
\end{document}